# $e^+e^-$ pairs from a nuclear transition signaling an elusive light neutral boson

A. Krasznahorkay*, F.W.N. de Boer†, M. Csatlós*, L. Csige*, Z. Gácsi*, J. Gulyás*, M. Hunyadi*, T.J. Ketel**, J. van Klinken‡, A. Krasznahorkay Jr.* and A. Vitéz*

*Institute of Nuclear Research of the Hungarian Academy of Sciences, P.O. Box 51, H-4001 Debrecen, Hungary
†National Instituut voor Kernfysica en Hoge-Energie Fysica NL-1098 SJ Amsterdam, The Netherlands
**Department of Physics and Astronomy of the Free University, de Boelelaan 1081, 1081 HV Amsterdam, The Netherlands
‡Kernfysisch Versneller Instituut, 9747 AA Groningen, The Netherlands

**Abstract.** Electron-positron pairs have been observed in the 10.95-MeV $0^- \to 0^+$ decay in $^{16}$O. The branching ratio of the $e^+e^-$ pairs compared to the 3.84-MeV $0^- \to 2^+$ $\gamma$ decay of the level is deduced to be $20(5) \times 10^{-5}$. This magnetic monopole (M0) transition cannot proceed by $\gamma$-ray decay and is, to first order, forbidden for internal pair creation. However, the transition may also proceed by the emission of a light neutral $0^-$ or $1^+$ boson. Indeed, we do observe a sharp peak in the $e^+e^-$ angular correlation with all the characteristics belonging to the intermediate emission of such a boson with an invariant mass of 8.5(5) MeV/c$^2$. It may play a role in the current quest for light dark matter in the universe.

## INTRODUCTION

Internal pair creation (IPC) can be instrumental in assigning spins and parities to nuclear levels. It was described with QED [1] long time ago. When an excited nuclear level does not decay by $\gamma$ radiation or by conventional IPC, an emission of a light neutral particle which in turn decays into an $e^+e^-$ pair is still a possible form of the deexcitation.

Such a light boson might be identified with a light and very weakly coupled neutral spin-1 gauge boson $U$, introduced by Fayet [2] more than two decades ago and revisited by Boehm and Fayet [3]. It was first argued by Boehm et al. [4] and recently by Fayet [5] and Beacom et al. [6] that light dark matter particles with masses between 1 and 20 MeV/c$^2$ annihilating through such bosons into $e^+e^-$ pairs may be the source of the observed 511 keV emission line in the galactic bulge [7]. They find that such a scenario is consistent with the observed dark matter relic density and other constraints from astrophysics and from particle physics.

In 1988, de Boer and van Dantzig [8] analyzed the emulsion results produced in relativistic heavy ion bombardments, an emulsion study, in which a distinct cluster of $e^+e^-$ pairs was observed at a short distance from the interaction vertex. These events were attributed to the emission and subsequent decay of a light neutral boson with a mass of around 9 MeV/$c^2$ and lifetime of about $10^{-15}$ s, remarkably occurring within a still allowed mass lifetime window of between 5 and 20 MeV/c$^2$ and $\tau \leq 10^{-13}$ s [9].

This finding motivated a systematic investigation for such a light neutral boson in searches for anomalous IPC in the decay of highly energetic nuclear levels in $^8$Be and $^{12}$C [10, 11].

The deviations between the measured (with poor angular resolution) and predicted IPC correlations of $e^+e^-$ in isoscalar magnetic transitions together with an apparent agreement with a corresponding E1 transition are explained as a consequence of the formation and decay of a short lived isoscalar neutral boson with spin-parity (pseudoscalar) $0^-$ or (axialvector) $1^+$ and a mass of about 9 MeV/$c^2$ [12].

The aim of the present work is to study a $0^- \rightarrow 0^+$ transition with better angular resolution in search of a light pseudoscalar boson. The electromagnetic transition between two nuclear states with zero angular momentum and opposite parities is forbidden and can occur by the emission of an e$^+$e$^-$ pair only if there is a parity admixture in the initial and/or final states or if there exists other than electromagnetic coupling between nucleons and the pair field [13, 14, 15, 16]. Because the light boson is a pseudoscalar object it should behave as a "magnetic" photon. Thus the allowed values of angular momentum and parity carried away by the boson in such nuclear deexcitation are $0^-, 1^+, 2^-$, etc.. The possibility of the existence of such 9 MeV/c$^2$ boson provided the motivation to investigate the 10.95 MeV $0^- \rightarrow 0^+$ g.s. transition in $^{16}$O.

Eklund and Bent [16] and Alburger [17] searched for $0^- \rightarrow 0^+$ transition by measuring the e$^+$e$^-$ decay of the 10.95 MeV state but a really sensitive search for those transition remained to be made.

## EXPERIMENTAL METHODS

In order to populate strongly and selectively the $0^-$ state, we used the $^{14}$N($^3$He,p)$^{16}$O reaction at E($^3$He) = 2.4 MeV [18], and performed the experiment in coincidence with the proton group populating the $0^-$ state. TaN target made at the KVI with an approximate thickness of 4 mg/cm$^2$ on a 5 $\mu$m thick Ta backing was used, allowing a beam current of 0.5 $\mu$A without considerable target damage during the course of the experiment. The Ta backing served also as an absorber for the $^3$He beam particles.

A plastic scintillator disk with a diameter of 45 mm and a thickness of 6 mm was used behind the target to get large solid angle and high count-rate capabilities for detecting the protons. A solid angle of about 2 sr and an energy resolution of about 1 MeV was achieved with such a detector for protons that populate the $0^-$ state.

For high-statistics measurements of the e$^+$ e$^-$ pairs, $\Delta$E - E detector telescopes from the IKF spectrometer constructed by Frölich at al. [11] with large solid angles were used. Three smaller ( $\Delta$E detector: 22 x 22 x 1 mm$^3$ and E detector: 30 x 30 x 70 mm$^3$) and two larger ( $\Delta$E detector: 38 x 40 x 1 mm$^3$ and E detector: 80 x 60 x 70 mm$^3$) telescopes were used at fixed angles. The target was inclined by an angles of 45 degrees with respect to the beam direction.

The telescope detectors were placed outside the vacuum chamber. To minimize the amount of the material around the target, a 24 cm long electrically conducting carbon fiber tube with a radius of 3.5 cm and a wall thickness of 0.8 mm, as well as thin

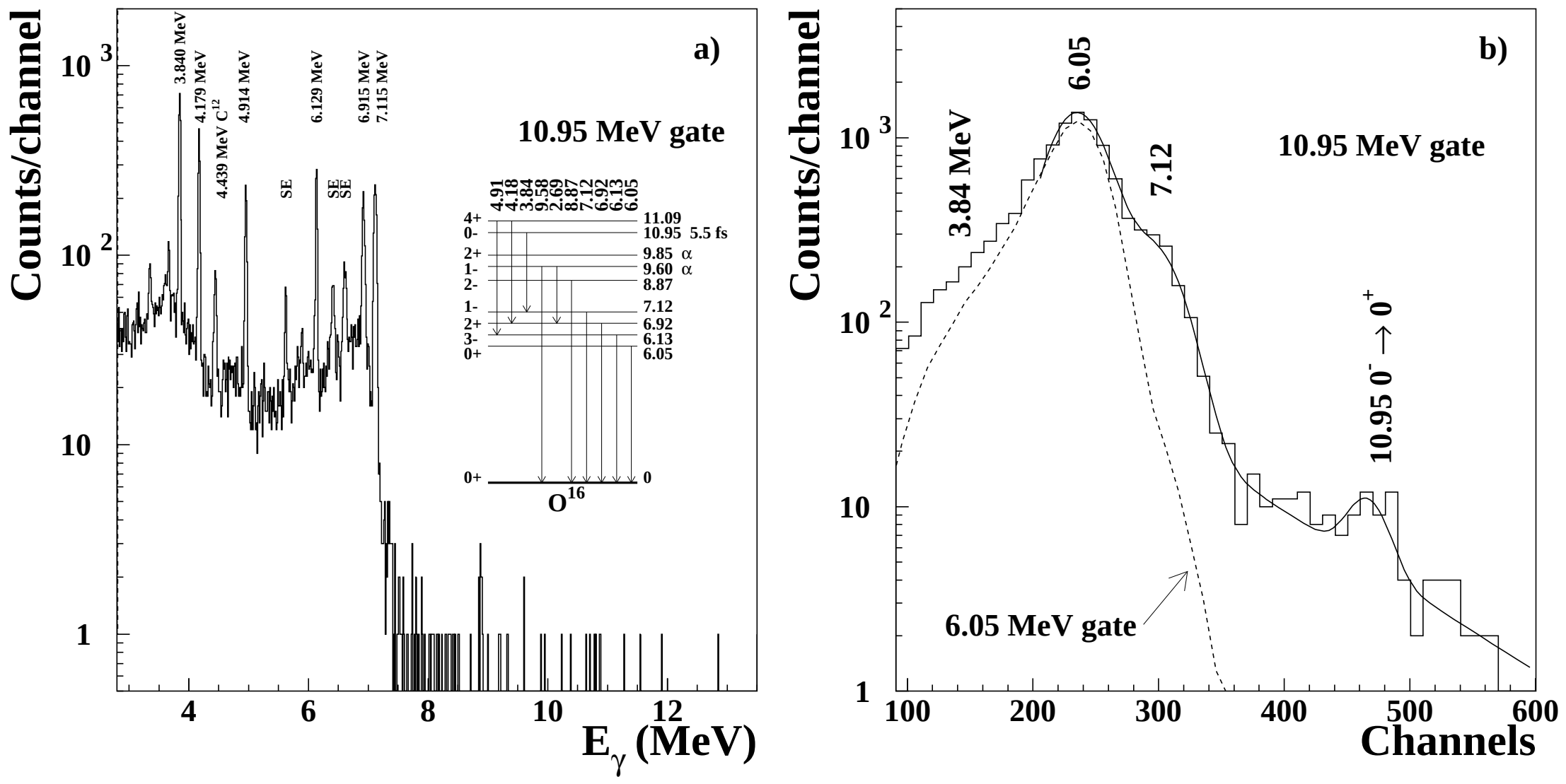

**FIGURE 1.** Internal pair creation and $\gamma$-ray spectra obtained from the $^{14}$N($^{3}$He,p)$^{16}$O reaction. See text for details.

aluminum target frames held by thin rods were used.

The $\gamma$-rays were also measured in coincidence with the protons. A highly efficient (120 %) clover type Ge detector equipped with a BGO anticoincidence shield [19] was used perpendicular to the beam and at a distance of 25 cm from the target.

## RESULTS

Fig. 1. shows an $e^+ + e^-$ sum spectrum and a $\gamma$-ray spectrum measured in coincidence with the protons populating the 10.95-MeV $0^-$ state. The 3.84-MeV $\gamma$-ray transition as well as a less intense $e^+$ - $e^-$ line at 10.95 MeV depopulating the $0^-$ state can be clearly seen, while no $\gamma$-rays could be observed at 10.95 MeV.

The energy calibration of the setup was performed with pairs of the 6.05-MeV transition and of the 4.44-MeV and 15.1-MeV transitions measured in the $^{10}$B($^{3}$He,p)$^{12}$C reaction with the same setup. The insert in the $\gamma$ spectrum shows a level scheme of $^{16}$O around 11 MeV, presenting the relevant $\gamma$ transitions, which are also indicated in the $\gamma$ spectrum.

Although, the energy resolution in the $e^+ + e^-$ sum spectrum is much worse ( FWHM$\approx$ 1.2 MeV) compared to the $\gamma$ spectrum, one can clearly observe a peak at 10.95 MeV, which corresponds to the $0^- \to 0^+$ transition in $^{16}$O.

The $\gamma$-ray background in the E detectors originating from the target is very well suppressed by requiring $\Delta$E - E coincidences in addition to the coincidence between the two telescopes. In $^{16}$O the 3.84-MeV (M1) and 7.12-MeV (E1) $\gamma$-$\gamma$ cascade depopulating the 10.95-MeV state may produce high energy electrons by Compton scattering in both $\Delta$E detectors. These events will produce similar coincidence signals as the $e^+e^-$ pairs.

Their energy sums add up to a broad spectrum with a sharp edge at 10.95 MeV and thus simulate the IPC process.

In order to estimate the contribution of such kind of background, measurements were performed by using $\Delta$E detectors with 3-mm thicknesses instead of the 1-mm ones. With the thicker $\Delta$E detectors we expected 9 times larger contributions from Compton-Compton coincidences. The results of these measurements confirmed our hypothesis that Compton-Compton coincidences contribute to the background. This kind of background drops at about 1 MeV above the $e^+e^-$ peak allowing a reliable background estimation.

As shown in Fig. 1, the $e^+e^-$ spectrum was fitted with an exponentially falling background and using 3 Gaussians corresponding to the 6.05-MeV transition, the peaks around 7 MeV and the 10.95-MeV $e^+e^-$-line. The intensity ratio obtained for the 10.95 MeV and 7-MeV lines is: $R_e = I(10.95\text{ MeV})/I(7\text{ MeV}) = 0.045 \pm 0.012$.

The 7-MeV lines contain the 7.12-MeV transition, which perfectly measures the excitation of the 10.95- MeV state as it decays by 100 % branching ratio to that level, and also the 6.92-MeV, $2^+ \rightarrow$ g.s. transition.

The intensity ratio of the 6.92 and the 7.12-MeV gamma lines was deduced from the gamma spectrum: $R = I(7.12)/I(6.92) = 1.20$. And the branching ratio, $B$, of the $e^+e^-$ decay was obtained for the $0^-$ state as:

$$B = \frac{R_e}{R}\alpha(6.92) + R_e\alpha(7.12) = (2 \pm 0.5) \times 10^{-4}, \tag{1}$$

where $\alpha(7.12)$ and $\alpha(6.92)$ denote the theoretical internal pair creation coefficients [20] of $2.56\times10^{-3}$ and $1.96\times10^{-3}$ for the 7.12-MeV E1 and the 6.92-MeV E2 transitions, respectively. As the lifetime of the $0^-$ state is known [21] to be $8 \pm 5$ fs, the partial lifetime of the $e^+e^-$ transition will be $(4 \pm 2) \times 10^{-11}$ s. This lifetime is comparable to the lifetime of a second-order electromagnetic process and in this sense agrees with the estimate for a pseudoscalar boson.

Assume that a short-lived boson with mass $m_u$ is created in such a nuclear transition with transition energy $E$ and decays into an $e^+e^-$ pair. In the center of mass system of the boson the $e^+$ and $e^-$ are emitted under 180°. In the laboratory system the opening angle will be smaller than 180° and depend on the $e^+$ and $e^-$ energies. By assuming an isotropic emission of the $e^+e^-$ pairs in the c.m. system, the angular correlation of the pairs can be calculated. The correlation has always a sharp maximum ($FWHM \leq 2°$) and the invariant mass $m_x$ is determined by the scattering angle $\Theta$, measured in the laboratory system as follows [12]:

$$m_u^2 \approx (1 - y^2)E^2 sin^2(\Theta/2), \tag{2}$$

where $E = E^+ + E^- + 1.022$ MeV is the transition energy, $E^{+(-)}$ the kinetic energy of the positron (electron) and $y = (E^+ - E^-)/(E^+ + E^-)$ is the disparity, all in the laboratory system. As an example, for the 10.95-MeV transition and a 9-MeV/$c^2$ boson we find a correlation angle, $\Theta$, of about 111°.

By measuring the above angular correlation pattern of the $e^+e^-$ pairs the mass of the new boson can in principle be deduced. However, the angular resolution of the IKF spectrometer is poor ($\approx 50°$ due to the choice of the large solid angles of the telescopes aiming at high statistics.

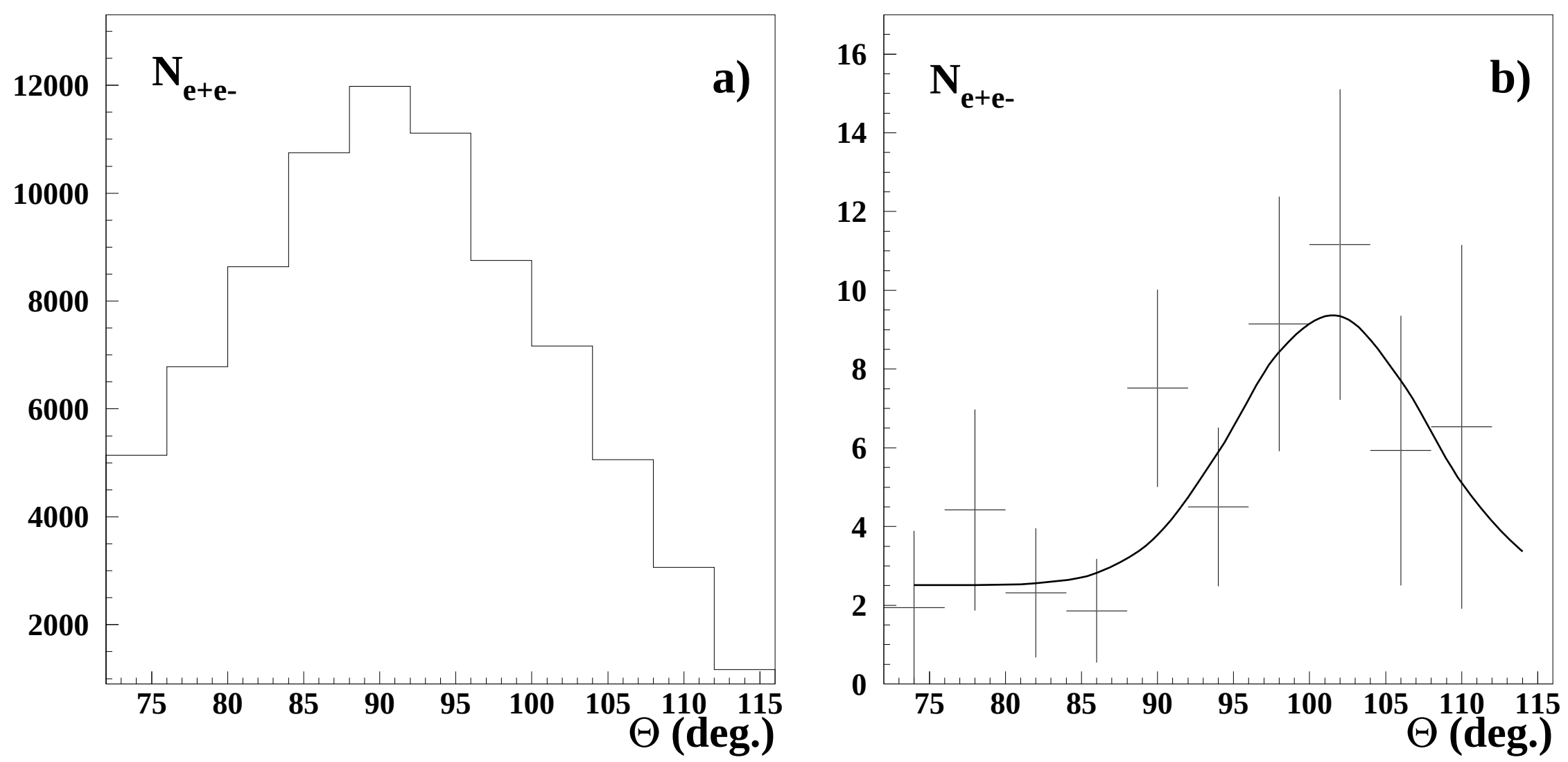

**FIGURE 2.** The angular correlation of the $e^+e^-$ pairs obtained from the decay of the 6.05-MeV (a) and from the 10.95-MeV (b) transitions in $^{16}O$.

The experimental angular correlation obtained with this setup resembles an E0 angular correlation (calculated with the formulas published by Rose [1]). An M0 transition may proceed through some higher order effects and may produce similar angular correlation pattern to an E0 transition.

In order to increase our sensitivity for searching the boson, we improved the angular resolution by inserting a standard MWPC detector between the ΔE and E detectors of the biggest telescopes.

The x,y position of the hits could be measured with an accuracy of 5 mm, which results in an angular resolution with FWHM=8.5° between the electrons and positrons in the 60 - 120° angular range. The actual angular resolution of the telescopes was even worse because of the scatterings on the plastic ΔE detectors.

Fig. 2 a) and b) shows the results of the angular correlation measurements obtained for the 6.05-MeV and for the 10.95-MeV transitions, respectively. As the angular correlation for the 6.05-MeV E0 transition is varying smoothly (almost linearly) in the investigated angular range Fig. 2a) shows the effective solid angle of the telescopes as a function of the correlation angles. In Fig. 2b) the counts have already been corrected for the the effective solid angle of the telescopes determined from Fig. 2a).

As the angular correlation pattern expected from the decay of a boson is sharper than the angular resolution of our telescopes, we expect a Gaussian like distribution with a width determined by the experimantal angular resolution. Using a Gaussian with free parameters, we fit the experimental angular correlation. The result is shown in Fig. 2b). with full line. We obtained $\Theta = 102.5^\circ \pm 2.0^\circ$ from the fit, which gives $m_u = 8.5 \pm 0.5$ MeV/c$^2$ for the mass of the boson. This uncertainty contains statistical uncertainties only.

## CONCLUSION

We observed for the first time the 10.95-MeV $0^- \to 0^+$ decay in $^{16}$O by observing $e^+e^-$ pairs. The energy sum of the pairs corresponds to the energy of the transition (10.95 MeV) and the observed branching ratio, $B = (20 \pm 5) \times 10^{-5}$, agrees with the expected boson branching ratio. The angular correlation of the pairs can be described by assuming the decay of a boson with mass of $m_u = 8.9 \pm 0.2$ MeV/c$^2$.

## ACKNOWLEDGEMENTS

The authors acknowledge H. Fraiquin (KVI, Groningen) for making the TaN target and the Van de Graaff generator staff at Atomki for their support during the course of the experiment. This work has been supported by the Hungarian OTKA Foundation No. T038404.